\documentclass[default,10pt]{sn-jnl}

\jyear{2021}%

\theoremstyle{thmstyleone}%
\theoremstyle{thmstyletwo}%

\theoremstyle{thmstylethree}%

\begin{document}

\title[The Relational Interpretations on soft matter as intermediate asymptoitcs]{The Relational Interpretations on soft matter as intermediate asymptoitcs}


\author*[1]{\fnm{Hirokazu} \sur{Maruoka}}\email{hmaruoka1987@gmail.com}\email{hirokazu.maruoka@yukawa.kyoto-u.ac.jp}

\affil*[1]{\orgdiv{Advanced Statistical Dynamics}, \orgname{Yukawa Institute for Theoretical Physics, Kyoto University}, \orgaddress{\street{Kitashirakawa Oiwakecho}, \city{Sakyo-ku}, \postcode{606-8502 }, \state{Kyoto}, \country{Japan}}}


\abstract{In this paper, it is demonstrated that there is a parallelism between the relational interpretation of Rovelli and the interpretation of soft matter based on intermediate asymptotics. The general interpretation of physics strongly assumes the duality of the observer and the world, and the uniqueness of the world though the relational interpretation suggested different conclusions: {\it no properties, no interaction}, and {\it facts are relative}. These conclusions are seemingly counterintuitive, though this work shows that similar conclusions are found in the interpretation of soft matter based on the concept of intermediate asymptotics. The interpretation of soft matter based on intermediate asymptotics also concludes that the properties are not determined without the scale. This is due to the conclusion of intermediate asymptotics that any formalization and its interpretation are localized by the scale. It is demonstrated that the similarity between the two interpretations originated from its monism of relations. This logical structure is also compared with the works in other disciplines. This work reports the insight that the relational interpretation can be a general and fundamental concept, not the one applicable to special cases.}

\maketitle 

\section{Introduction}\label{sec1}

Physics is the most fundamental scientific discipline concerned with the nature and properties of matter and energy\cite{MaxwellMat, FeynmanLec1}. Originating from the Greek word for 'nature', it has been the study of nature, namely the world surrounding us. Physics seeks universal properties about nature, the properties that are independent of time and space. This strong motivation for universality is an outstanding character of physics, and it has become the basis of application sciences including chemistry, and engineering.  

The study of nature naturally leads to the objectivization of nature. Nature has been dehumanized and mechanized\cite{PrigogineOrder}. This process is based on the assumption of the distinguishability of subject and nature. Here a fundamental duality, the subject and the object, observers and the world can be found. If the world is distinguishable from the observer and is isolated, and if the world we investigate is unique, then the properties of that world should be unique as well. This leads to the concept of the uniqueness of the world. These assumptions of the duality and uniqueness of the world constitute the general interpretation of physics.

However, a fundamentally different interpretation is proposed by Rovelli\cite{RovelliOT,RovelliHe,RovelliAr}. His theory of the Relational Interpretation suggested the interpretation that is beginning from their {\it relations}. Relations are interactions connecting physical objects. The relations constitute the network in which objects and subjects are equally involved. This leads to counter-intuitive conclusions that are fundamentally different from the general interpretation of physics though this interpretation provides a consistent view on quantum mechanics. Rovelli developed his idea in quantum mechanics though the essential idea with regard to relations can be general.

In this paper, the author attempts to support the relational interpretation and to demonstrate that there is an interpretation similar to Rovelli's relational interpretation in soft matter physics. In the field of soft matter, one frequently observes the phenomena in which its property changes depending on the {\it scale} of the physical parameters. Here we found that the {\it scale} plays the same role as {\it relations} do in the relational interpretation. This interpretation can be formalized by {\it intermediate asymptotics} proposed by Barenblatt\cite{Barenblatt1996,Barenblatt2003}. 

The relational interpretation in the soft matter also reveals the same conclusions as the relational interpretation in quantum mechanics. The property fundamentally depends on the scale in the same way as it depends on the relations. In particular, when the parameters related to subjects are involved, it also leads to the conclusion that facts are fundamentally relative.

I will demonstrate that such a similarity between the Relational interpretation on the soft matter and that of quantum mechanics is derived from the fundamental monism of relations in the relational interpretation, which is different from the dualism in the general interpretation of physics. 

\section{The general interpretation of physics}

Physics is a scientific discipline in which one explores the universality and laws of nature, namely the world. The study of nature has its origins from Greek philosophy though it has undergone great variations up to this day. Its methodology and its style of interpretation have been gradually established. Nature in physics became dehumanized, isolated objects that were supposed to be independent of subjects. The method of physics spontaneously became the established manner of objectivization of nature in which they consider nature as independent of humans, and its law as independent of space and time\cite{PrigogineOrderI}.

Throughout the history of the development of physics, we can find its strategy and that there are some assumptions for the general interpretation of physics. They can be characterized as follows, "the duality of the subject from the object", and "uniqueness of the world".  

\begin{itemize}

\item Assumption 1. {\it The duality of subject and object that are independent of each other} : it means that physics always starts from the fundamental duality which are independent, distinguishable from each other; the subject and the object, the observer and the world. One can separate the subjective domain from the objective domain without changing their properties because two elements are independent entities in principle. Therefore, the proposition of physics can be exemplified by the influence of subjects as far as it is concerned about only objects. 

\item Assumption 2. {\it The uniqueness of the world} : the world is unique, therefore we can explore universality. Due to the first assumption, the subjective domain and the objective domain are distinguishable without losing their entities.  The objective domain is free from individuality and constitutes the world surrounding other subjects. If we eliminated the subjective matters, the description of the world should converge uniquely and the laws or knowledge obtained from local experimentation can be globally applicable. Therefore, the propositions of the objective domain possess universality. 

\end{itemize}

The isolation and the objectivization of nature require the assumption 1. The isolation of nature corresponds to the elimination of subjective matters, the human. The exploration of the universality requires the assumption 2. Experimentation provides knowledge of the unique world surrounding us. Therefore, the knowledge of local experimentation can be the global one, which is applicable for other subjects. It makes scientists believe that the properties of matter potentially exist without human, observers. The existence of human beings or subjects is not essential for nature and physics\cite{PrigogineOrderII}.

These two assumptions are essential for exploring universality. It seems impossible to explore universality without them. It does not matter whether these assumptions are conscious or unconscious for scientists. These two assumptions are conditions of possibility for exploring the universality. The enterprise of physics in which they pursue the global truth has already been possible by assuming the duality and the uniqueness. It is the general style of interpretation of physics. Scientists attempt to interpret nature based on this interpretation consisting of these assumptions.

\section{Rovelli's relational interpretation}

While the general interpretation of physics is based on duality and uniqueness which makes scientists have the interpretation that the world is positive, independent of human being and unique. Rovelli, however, proposed a new interpretation for physics\cite{RovelliOT,RovelliHe,RovelliAr}. This new approach had been elaborated while he was pursuing the origin of time, a consistent interpretation of quantum physics. Since Rovelli's relational interpretation was developed in quantum mechanics, it is heavily based on the formalization of quantum mechanics. However, its essential idea is simple and its key concept is {\it relations}. Here I call the relational interpretation the interpretation based on the relation, which was mainly summarized and described in his book\cite{RovelliHe}. Then I demonstrate that the similar structure can be found in the theory of intermediate asymptotics in soft matter even if it belongs to classical mechanics.

Here I define that the essential idea of the relational interpretation is {\it to start with relations for interpreration of physics}. Here relations mean any interactions that connect between objects. The important point is that this idea is applied to subjects or observers as well. The relational interpretation considers subjects or observers as an involved part of nature. He emphasized that "the simple observation that scientists as well, and their measurement instruments, are all part of nature\cite{RovHe1}." Starting from relations corresponds to starting from a perspective in which the subject is part of nature. Then the world is a network of relations and objects. This idea leads to the following radical conclusions\cite{RelInt}.

\begin{itemize}

\item Conclusion 1. {\it There are no properties outside of relations}. It insists that any properties do not exist without interactions. Here the properties of an object are the way in which it acts upon other objects. Unrelated objects and properties do not simply mean unawareness but nonexistence. Generally, we believe that the property potentially is attributed when it does not interact. However, the relational interpretation concludes such a notion is superfluous and misleading. 

\item Conclusion 2. {\it Facts are relative}. As the first conclusion said, there are no properties outside of relations. This immediately means that it is possible that a property manifests itself for a related observer but it does not for another unrelated observer. Property is relation-dependent, which means that facts for observers are also relative.  

\end{itemize}

Rovelli exercised his idea in quantum mechanics and he insists that quantum mechanics describes the way in which one part of nature manifests itself to any other single part of nature. He treats the state of a quantum system as being observer-dependent. The state is the relation between the observer and the system. This interpretation succeeded in eliminating the paradox of quantum theory. For example, the relational interpretation shows that the apparent incongruity on {\it entanglement} raised by what seemed like communication at a distance between two entangled objects was due to neglecting the existence of a third object that interacts with both systems. This paradox stems from our dogma that the properties of objects potentially exist even when they do not interact.

Comparing the general interpretation of physics, it is easily found that the fundamental difference is the form of its origin. The general interpretation of physics always starts from the dualism of subject and object, observer and nature while the relational interpretation is the monism of relations though relations are the origin of the duality. Relations create elements of the dualism of subjects and objects. This monism simply means the fundamental indistinguishability of subjects and objects. Therefore, this dualism can be seen as relation-dependent and it is relative. Therefore, the second assumption is simply relativized by introducing a relation. If we say that the world is a network of relations and objects, the world is no longer unique for {\it any} objects but it is unique within {\it their own} network. This is not a contradiction.

Note that the subject in the relational interpretation is different from the subjects in any other interpretation of physics. QBism (Quantum-Bayesianism) also introduces the subject to describe the world in the limited information given to the subject\cite{QBism}. However, the subject of QBism is fundamentally different from the subject in relational interpretation in the point that QBism considers the subject as an entity independent of the objects, and the world. It also starts from the dualism of two elements independent of each other. The subject interacts with the world through the information but this interaction never changes how two elements are. However, the subjects in the relational interpretation reveal a fundamental dependence on others. How it manifests to others can vary depending on the relation. The subjects are no longer in privileged positions that are exempted from the interference of others but are fully involved in the network. This subjectivity indistinguishable from objects is essential on the subject of relational interpretation. 

In the next section, I will show this based on the framework of intermediate asymptotics. In the framework, the {\it scale} plays the same role as the {\it relation} in the relational interpretation, which is manifested by a dimensionless number ${\rm De}$. The framework is provided by the concept of intermediate asymptotics.

\section{Intermediate asymptotics and the scale as relation}

The idea of Rovelli's Relational interpretation is to start from relations. As a result, properties are relative in the sense that they can be different in different network of relations but it is unique in the sense that it is uniquely defined within the network of their relation. Therefore, the uniqueness is {\it localized by relation} in the relational interpretation while in the general interpretation, the uniqueness is global, universal for any observers. 

The property is localized by the relation. In fact, I found that such a logical structure can be found in the concept of intermediate asymptotics as well.

An {\it intermediate asymptotic} is an asymptotic solution valid in a certain scale range, which has been introduced and formalized by Barenblatt\cite{Barenblatt1972,Barenblatt1996,Barenblatt2003,Barenblatt2014,Goldenfeld,Goldenfeld1992,Oono2013,Maruoka2023}. It can be considered as the formalization of a physical model and its idealization.

Suppose an arbitrary physical function,
\begin{equation}
y =f\left(x,  z, t \right)
\label{eq:ES1c}
\end{equation}
of which $y, x,z, t$ are physical parameters having certain dimensions. Any physical functions such as Eq.~\ref{eq:ES1c} can be transformed to a following self-similar solution,
\begin{equation}
\Pi = \Phi \left( \eta,  \xi \right)
\label{eq:ES1a}
\end{equation}
where $\Pi = y/t^{\alpha} $, $\eta = x/t^{\beta}, \xi = z/t^{\gamma}$. Their power exponents $\alpha, \beta, \gamma$ are determined by dimensional analysis or consideration of invariance\cite{SS}. According to the recipe by Barenblatt\cite{Recipe}, nextly one considers the convergence of $\Phi$. If $\Phi$ converges to a finite limit as $\xi$ goes to zero or infinity, $\eta$ can be excluded from consideration. For example, if $\Phi \rightarrow {\rm const}$ as $\xi \ll 1$ then we have an following intermediate asymptotics,
\begin{equation}
\Pi = \Phi \left( \eta \right)~\left(  \xi \ll 1 \right)
\label{eq:ES1b}
\end{equation}
while Eq.~\ref{eq:ES1b} is valid in the range for $\xi \ll 1$ or $0 < z \ll t^{\delta}$\cite{Int}. 

Eq.~\ref{eq:ES1b} is an intermediate asymptotics as it is valid in a certain scale range for $\xi \ll 1$.  If $\eta$ also satisfy the same condition, $\Phi \rightarrow {\rm const}$ as $\eta \ll 1$, then $\Pi = {\rm const}$. Thus we have a following intermediate asymptotics, $y = {\rm const}~ t^{\alpha}~\left(\eta \ll 1,  0 < x \ll t^{\beta} \right)$. This is a power-law valid in a local scale. It is also an intermediate asymptotics.

Functions of which parameters consist of power law monomials are called self-similar solutions as they are invariant for scale transformations such as $x \rightarrow A^{\alpha}x~\left(\alpha > 0 \right)$. Their parameters, $\eta$, $\xi$ are called similarity parameters. Any physical functions have physical parameters with dimensions. There is at least one self-similar form invariant for scale-change of dimensions.

As has been shown, intermediate asymptotic is an asymptotic expression valid in a local scale range in which similarity variables go to 0 or infinity. 

\begin{figure}[h!]
\centering
\includegraphics[width=1\columnwidth]{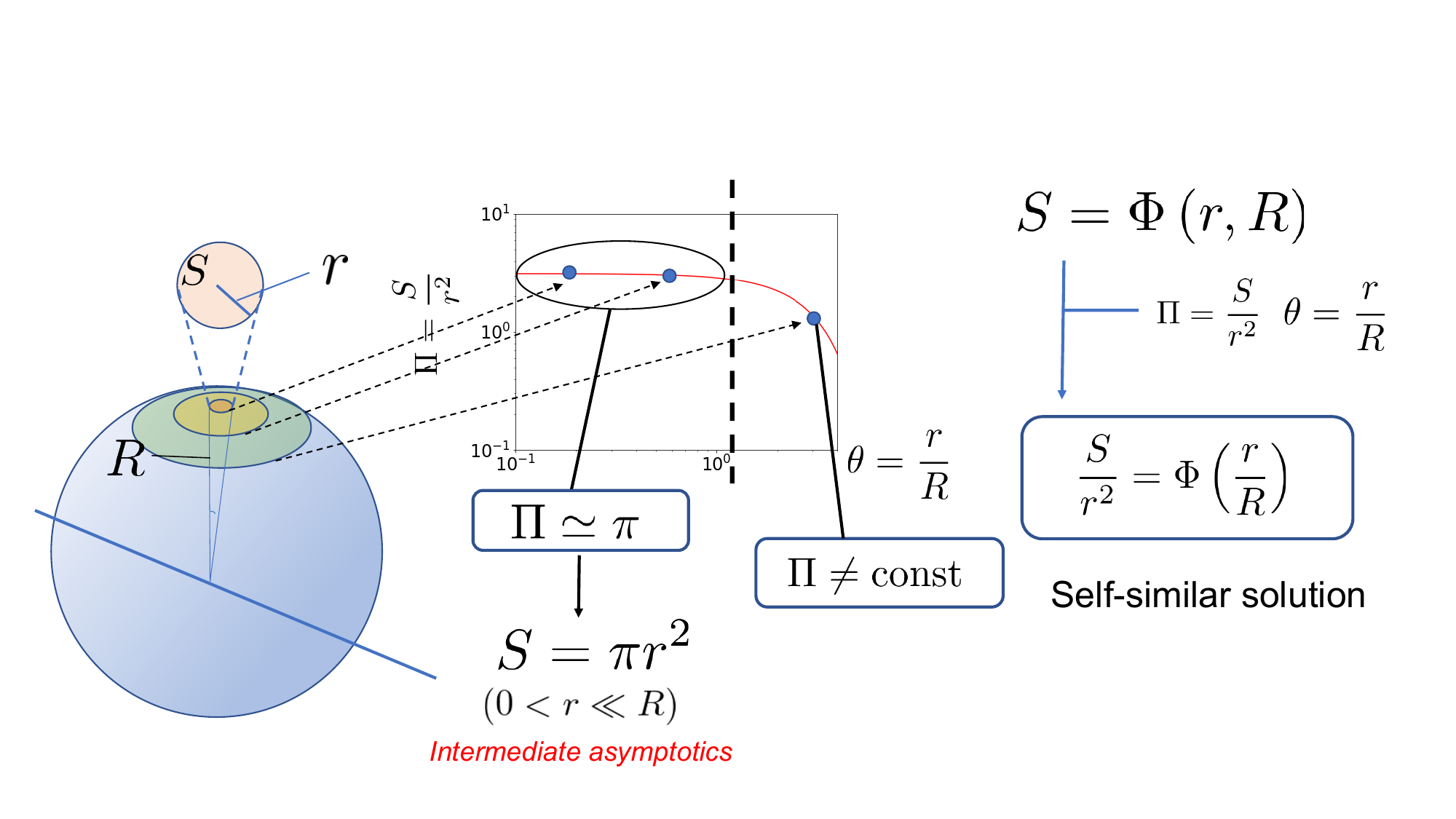}
\caption{A circle of which radius is $r$ and surface area is $S$, is illustrated on a sphere of which radius is $R$. There is a self-similar function $\Pi =\Phi \left( \theta \right)$ where $\Pi = S/r^2$ and $\theta = r/R$ to connect these parameters. The plot between $\Pi$ and $\theta$ suggests, the dimensionless number $\Pi$ is not constant in all scale range of parameter $r$,  but dependent on $\theta$. However, there is a scale range in which $\Phi$ converges to a finite limit,$\Phi \rightarrow \pi$ as $\theta \rightarrow 0$, namely $\theta \ll 1$. In this scale range, there is a scaling law valid in a certain scale range as $S=\pi r^2~\left(0 < r \ll R \right)$, which corresponds to an {\it intermediate asymptotics}.  \label{fig:FS1a}  }
\end{figure}

To demonstrate this concept, the following example is helpful. Imagine that a circle is pictured on the surface of a sphere (See Fig.~\ref{fig:FS1a}). In this problem, the involved physical parameters are the surface area of circle $S$, radius of the circle $r$ and the radius of sphere $R$. Here we would like to know the scaling behavior between $S$ and $r$. Therefore we assume the functional relation as follows: $S = \Phi(r, R)$. 
In this case, we attempt to obtain the exact scaling behavior by dimensional consideration. According to dimensional analysis, as the dimension of physical parameters $[S]=L^2$, $[r]=L$ and $[R]=L$, selecting $r$ as a governing parameter of independent dimension, we have the following self-similar function,
\begin{equation}
\Pi = \Phi \left( \theta \right)
\label{eq:ES1}
\end{equation}
where $\Pi = \frac{S}{r^2}$ and $\theta = \frac{r}{R}$. Eq.~\ref{eq:ES1} suggests that we expect the following scaling relation, $S \sim r^2$, if $\Pi$ is constant. However, we easily find that this guess depends on the behavior of $\Phi$. 

By the geometrical consideration, we can calculate the exact form of $\Phi$ as follows,
\begin{equation}
\Phi\left(\theta\right) =2\pi \frac{1- {\rm cos} \theta}{\theta^2}.
\label{eq:ES2}
\end{equation}
To know the behavior of $\Phi$ in the case in which $\theta \rightarrow 0$, which corresponds to the increase of $R$ or the decrease of $r$, Taylor expansion is applied to Eq.~(\ref{eq:ES2}) then we have,
\begin{equation}
\Pi =\Phi\left(\theta\right) \simeq \pi -\frac{\pi}{12}\theta^2 \cdots +\underset{\theta \rightarrow 0}{\longrightarrow} \pi.
\label{eq:ES3}
\end{equation}
As Eq.~(\ref{eq:ES3}) shows, $\Phi$ converges to a finite limit $\pi$, then we have a following intermeidate asymptotics as $\Pi = \frac{S}{r^2}$,
\begin{equation}
S = \pi r^2~\left(0 < r \ll R\right)
\label{eq:ES4}
\end{equation}
as far as the asymptotic condition $\theta \ll 1$, corresponding to $0 < r \ll R$, is satisfied.

Note that the scaling law Eq.~(\ref{eq:ES4}) is valid in the scale range  $(0 < r \ll R)$, in which the circle is significantly smaller than the sphere. Therefore, Eq.~(\ref{eq:ES4}) is an asymptotic expression which is valid in the certain range of variable $r$. This scaling law formalized {\it locally} is an intermediate asymptotic in this problem.

Every physical problems can be transformed to self-similar solutions as far as they have dimensions. By considering the convergence of $\Phi$, essential similarity variables can be selected to have the idealized solution effectively and practically. The convergence of $\Phi$ can be verified by the experimental or simulational results. The exact form of $\Phi$ is not necessary for this procedure.

The important point of this concept is that this process, in which one screens the self-similar variables of $\Phi$ depending on their convergence, corresponds to the idealization of the problems. More or less, all the physical models involve idealizations such as ignorance of friction force, ignorance of quantum or relativity effect. All these assumptions correspond to the idealizing process of dimensionless function. For example, ideal gas equation can be considered as an intermediate asymptotic valid in the range where the volume of molecules $b$ and the molecular interaction $a$ are negligible on the van der Waals equation as follows,
\begin{equation}
p=\frac{nRT}{V-nb}-\frac{an^2}{V^2} \longrightarrow \frac{nRT}{V}~~\left(\frac{an^2}{V^2} \ll p \ll \frac{RT}{b} \right).
\label{eq:ES5}
\end{equation}
This idealizing scale range is satisfied as far as $\Pi_a = \frac{a n^2}{pV^2} \ll 1$ and $\Pi_b = \frac{pb}{RT} \ll 1$.

The concept of intermediate asymptotics suggests that actual problems are fundamentally {\it local} for the observers describing the problems. Therefore, the description is relative in the sense that it is locally valid for a certain observer. However, it is unique if the observer belongs to the same locality. It is easy to see that this logical structure is similar to that of the relational interpretation.

I will show that intermediate asymptotics reveal similarity with the relational interpretation if the similarity parameter includes the condition of the subject. If the intermediate asymptoitics is the representation localized by the condition of a certain subject, the phenomenon is localized by a certain subject.

\subsection{The pitch drop experiment} 

The terms of soft matter is introduced by Pierre-Gilles de Gennes, which denotes the materials characterized by complexity and flexibility\cite{deGennes1992,deGennesSC}. It deals with the polymer, the dispersed solution, the emulsion. In this field, one occasionally encounters the materials or phenomena which drastically change their behaviour depending on the scale of the physical parameters. The pitch drop experiment is a good example of this. I will show how the pitch drop experiment is an intermediate asymptotics localized by the time of observation.

The pitch drop experiment is an experiment to observe the long time-scale behavior of pitch\cite{Parnell}. It is performed by pouring the heated sample of the pitch into a sealed funnel, then cutting the neck of the funnel to let it flow. The solid pitch is so heavily viscous material that any flow of the pitch is not observed for months or years. For 75 years, however, seven drops were observed. This experiment was intended to teach that any solid looking materials can behave like a fluid on a larger scale of time. One of the studies won Ig-Nobel prize.

This experiment is occasionally treated as a trivia but it seems that it is posing a fundamental question of physics. In general, the physical properties, whether it is solid or fluid, are totally considered to be objective properties, which do not depend on time. However, the pitch drop experiment raises a question to this point. If one is asked a question whether the pitch is solid or fluid when months or years have passed since the pitch was prepared, one may answer that it is solid because any fluidic behavior is not objectively observed. However, if one is asked the same question after 100 years, the answer will be different. This seems contradictory to the fact that physical properties are objective. If one asks the question: Is the pitch fluid or solid? how can we answer this question? 

It seems that this experiment suggests that any physical materials and properties are "relative" on the time-scale, therefore we have to consider the material properties with the scale. Let us formalize this problem by intermediate asymptotics to see how the behavior of pitch varies depending on time.

Assuming that the flow through stem obeys Poiseuille's law, the rate of flow of the pitch out of the funnel is described as follows,
 \begin{equation}
\frac{d V}{d \tau} = \frac{\pi d^4 \rho g}{128 \mu}\left( 1 + \frac{h}{l}\right) 
\label{eq:E1} 
\end{equation} 
where $V$ is the volume of the pitch, $\tau$ is the time, $\rho$ is the density of pitch, $\mu$ is a viscous coefficient,  $d$, $l$ are the diameter and the length of the stem, $h$ is the depth of pitch in the funnel.

These parameters belong to the material itself. However, as the qualitative discussion shows, the property of the material varies depending on the time scale of the observation. Now let us introduce the time scale of observation $T$ and see how the behavior reveals depending on $T$, not $\tau$. It corresponds to seeing the phenomenon in different time-scales, and can be realized by the method of multiple scales\cite{Holmes}.

Assuming a different time scale of the observation $T_1 = \tau$, $T_2 = \varepsilon \tau$ where $\varepsilon = \frac{T_2}{\tau} \ll 1$. Here we have two different time scales: $T_1$ is the same time scale as the actual relaxation time, and $T_2$ is a much shorter time scale than its actual relaxation time. Assuming these time parameters, $T_1$ and $T_2$, we have
 \begin{equation}
\left( \frac{\partial }{\partial T_1} +\varepsilon \frac{\partial }{ \partial T_2} \right)V = \frac{\pi d^4 \rho g}{128 \mu}\left( 1 + \frac{h}{l}\right).
\label{eq:E2} 
\end{equation} 
From this, in the time scale $O\left(1 \right)$ we have
 \begin{equation}
\frac{\partial V}{\partial T_1} = \frac{\pi d^4 \rho g}{128 \mu}\left( 1 + \frac{h}{l}\right) 
\label{eq:E3} 
\end{equation} 
which corresponds to Eq.~(\ref{eq:E1}). However, in the scale of $O\left(\varepsilon \right)$, we have
 \begin{equation}
\frac{\partial V}{\partial T_2} = 0,
\label{eq:E4} 
\end{equation} 
which means that no flow is observed.

Here we introduce the Deborah number ${\rm De} = \frac{\tau}{T}$, which is defined as the ratio between the relaxation time $\tau$ and the observation time $T$.  In the time scale in which $T= T_2$ or ${\rm De} \gg 1$, which corresponds to the observation of the pitch in the time scale much smaller than the actual relaxation, the behavior is described in Eq. \ref{eq:E3}, which means that no flow is observed and it can be considered as the solid behavior. However, in the time scale of $T = T_1$ or ${\rm De} = 1$,  which corresponds to the observation of the pitch in the same time scale as the actual relaxation, the behavior is described by Eq. ~\ref{eq:E4}. It corresponds to Eq. \ref{eq:E1} and the behavior of fluidity reveals. Therefore the entire behavior can be described as follows,
\begin{eqnarray}
  \frac{dV}{dT} = \begin{cases}
    0 ~ \left( {\rm De} \gg 1 \right)   \\
    \frac{\pi d^4 \rho g}{128 \mu}\left( 1 + \frac{h}{l}\right)  ~ \left( {\rm De} \sim 1 \right).
  \end{cases} \label{eq:E5} 
\end{eqnarray}
These two asymptiotic exspressions {\it locally} defined by ${\rm De}$ are intermediate asymototics.

In this formulation, this result is somehow seemingly trivial for the multiple scale, since there is no term for $O\left( \varepsilon \right)$. There can be critics that there is no need to apply the multiple scale. However, we must remember that the lesson of the pitch drop experiment is how the property of the material changes depending on the time scale of the observation $T$. To describe this lesson, the observation time $T$ is necessary to consider and the application of the multiple scale is essential to describe the different behavior in different time scale. The consideration of $O\left( \varepsilon \right)$ is essential to show the solid behavior; there is no flow on the solid as $\frac{d V}{d T} = 0$. 

\subsection{Viscoelastic behaviors on the dynamical contact} 

A distinct properties which depends on the time scale can be seen not only in the pitch drop experiment but in the various phenomena, dealt in the area of rheology or soft matter physics. Silly Putty is a toy which entertains by its behavior depending on the time scale of the loading force. If one throws the Silly Putty against the wall, it will bounce like an elastic ball though it will flow and spread on the ground by settling on the floor. This is also exactly a similar case. If one throws the ball onto the wall, the loading force $T$ is applied in smaller time scale than relaxation time $\tau$, then Deborah number is ${\rm De} \gg 1$. If one settle it on the ground the time range of applying force $T$ is constantly applied then it decrease ${\rm De} $ to reveal the fluidic behavior. Recently the author described such a functional change for the dynamical impact on the dusted viscoelastic board\cite{Maruoka2023}. Here I show such variation of response arises from Maxwell viscoelastic model.

Maxwell model is the model for viscoelastic materials consisting of a unit in which the spring and the dash-pod are serially connected. The behavior of materials to which the certain deformation $\epsilon$ is applied can be described by the following differential equation, $\frac{\mu}{E}\frac{d \sigma}{dt} + \sigma = \mu \frac{d \epsilon}{dt}$ where $\mu$ is viscous coefficient originated from dash-pod and $E$ is elastic modulus derived from spring part. It was found that the variation of response arises when the material receives the constant rate of deformation as $\frac{d \epsilon}{dt} = {\rm const}$. In this case, the stress $\sigma$ exerted on the material is described as follows:
 \begin{equation}
\sigma =\mu \frac{d \epsilon}{dt} \left[ 1 - \exp \left(-\frac{E t_c}{ \mu} \right) \right] 
\label{eq:E6} 
\end{equation} 
where $t_c$ is contact time. Here an inverse Deborah number is $Z= 1/{\rm De}  = \frac{E t_c}{\mu}$. To visualize the transition of the character of the material, the following transformation takes place.

The solid behavior is here characterized by Hooke's law, which is expressed by the $\sigma = E \epsilon$. Here one can define a dimensionless number representing the Hookean behavior, namely solid behavior, as $\Pi = \frac{\sigma}{E \epsilon }$. Since constant deformation is here assumed, the contact time is estimated as $t_c = \epsilon / \frac{d \epsilon}{dt} $. Then $Z = \frac{E \epsilon}{\mu}  /\frac{d \epsilon}{dt} $. Thus, Eq. (\ref{eq:E1}) is expressed as:
 \begin{equation}
\Pi= \frac{1}{Z} \left[ 1 - \exp \left(-Z  \right) \right] .
\label{eq:E7} 
\end{equation} 
Here we have a self-similar solution $\Pi = \Phi\left(Z\right)$. 

To see the behavior of $Z \ll 1$ and the other, let us see the asymptotic behavior of Eq. (\ref{eq:E7}) by applying Taylor expansion as follows,
 \begin{equation}
\Pi=\Phi\left( Z \right) \simeq 1 - \frac{1}{2 }Z  \cdots \underset{Z \ll 1  }{\longrightarrow} 1.
\label{eq:E8} 
\end{equation} 
As the expansion of $\Phi$ shows, $\Phi$ converges to a finite limit $\Phi \rightarrow 1$ as $Z \ll 1$, thus $\Pi = \frac{\sigma}{E \epsilon} = 1$. Then the behavior is totally characterized by the hookean behavior in $Z \ll 1$. In case where $Z$ is not small enough, the different behavior starts to appear, as $\Pi = 1 - \frac{1}{2} Z $. Thus, depending on the $Z$, the following behaviors appear,
\begin{eqnarray}
  \sigma = \begin{cases}
    E \epsilon~ \left( Z \ll 1 \right)   \\
    E \epsilon - \frac{1}{2} E^2 \epsilon^2 / \mu\frac{d \epsilon}{dt}  ~ \left( Z \sim 1 \right).
  \end{cases} \label{eq:E9} 
\end{eqnarray}
These are intermediate asymptotics {\it localized} by $Z$ in the contact. 

Such a variation of response is observed in the dynamical impact of a solid sphere onto PDMS viscoelastic board and that scaling behavior also varied\cite{Maruoka2023}. Such variation of scaling law is called crossover of scaling law. The scaling law of $t_c = \epsilon / \frac{d \epsilon}{dt} $ was experimentally confirmed, and the author successfully explained this crossover of scaling law by the following self-similar solution,
 \begin{equation}
\Psi=\frac{2}{3 } \frac{Z}{\left[1 - \exp\left( - Z \right) \right]}
\label{eq:E10} 
\end{equation} 
where $\Psi = \frac{ \delta_m^3 E \phi}{ R^{2} h \rho v_i^2} $ and $Z= \frac{E \delta_m}{\mu v_i}$, and $\delta_m$ is a maximum deformation, $\phi$ is the fraction of contact, $\rho$ is the density of the impactor, $R$ is a radius of impactor of sphere, $h$ is the thickness of PDMS board, $v_i$ is the impact-velocity.

By applying the singular perturbation method to this self-similar solution, one has the following asymptotic solution,
\begin{equation}
\delta_m = \frac{E\rho h R^2}{54 \phi \mu^2 } + \left(\frac{E \rho^2 R^4 h^2 }{486 \phi^2 \mu^{3} }\right)^{\frac{1}{3}} v_i^{\frac{1}{3}} + \left( \frac{2 h R^2 \rho }{3 \phi E} \right)^{\frac{1}{3}} v_i^{\frac{2}{3}}
\label{eq:E11}
\end{equation}
as $\varepsilon = \frac{E R}{\mu v_i } \rightarrow 0$. As Eq.\ref{eq:E11} shows, the high impact-velocity and the impact of larger sphere increase $Z$, then the impact is elastic and $\delta_m \sim v_i^{2/3}$ is dominant. On the other hand, the low impact-velocity and the impact of smaller sphere decrease $Z$, then the impact is viscoelastic and $\delta_m \sim v_i^{1/3}$ is dominant. Therefore depending on $Z= \frac{E \delta_m}{\mu v_i}$, the scaling law changes as follows,
\begin{eqnarray}
  \delta_m \sim \begin{cases}
    v_i^{\frac{2}{3}}~ \left( Z \ll 1 \right) ~{\rm Elastic~impact}  \\
    v_i^{\frac{1}{3}}~ \left( Z  \sim 1 \right) ~{\rm Viscoelastic~ impact}.
  \end{cases} \label{eq:E12}
\end{eqnarray}
These two are two intermediate asymptotics localized by $Z$ in the dynamical contact of Maxwell viscoelastic materials. 

In this case, we can find that even scaling laws vary depending on the scale as {\it relation}.

This insight suggested that the property of material cannot be determined without consideration of the scale. Again we see that the property of the material, which we tend to consider as the objective property, was localized by the scale.

\section{The relational interpretation of soft matter as intermediate asymptotic}

Both two cases revealed that the physical properties cannot be determined without considering the scale. This scale is expressed by the dimensionless number ${\rm De}$. ${\rm De}$ measures the scale.

In the pitch drop experiment, the subject is clearly introduced as the observation time $T$. $T$ is essential to consider how the property of the pitch changes depending on the scale of the observation time. Since the essence of the lesson of the pitch drop experiment was how the observation time scale changes the property, the introduction of $T$ into the problem was essential. However, one must note that it is the ratio of the relaxation time $\tau$ and the observation time $T$ that forms its relation on the pitch drop experiment(see Eq.~\ref{eq:E13}). ${\rm De}$ is the relation of the pitch. The reason why the pitch drop experiment seems to reveal the bizarre behavior is that simply it has the larger, but not too large relaxation time $\tau$ to observe its fluidity. Fundamentally, all materials should posses the same properties. The glass has the relaxation time of a hundreds thousand years\cite{Welch}. Conversely, the fluid of water reveals the solid behavior by an extremely instantaneous load of force. The soft materials are nothing but this intermediate behavior is apparent for {\it our} familiar scale. ${\rm De}$ provides different intermediate asymptotiics as in Eq.~\ref{eq:E5},
 \begin{equation}
{\rm De}= \frac{\tau}{T} = \frac{{\rm relaxation~time} }{{\rm observation~time}}. 
\label{eq:E13} 
\end{equation} 

In the viscoelasticity on the contact, it is different from the pitch drop experiments in the point that the scale does not act in a passive way like the observation. However, in this case, the scale appears as the contact time $t_c$.  $t_c$ is the active parameter in the sense that the contact itself is the active process (for the impactor). However, in this case, the scale of $t_c$ changes the response properties of the material as it was seen in the pitch drop experiment for $T$. When $t_c$ is so short, the behavior is totally elastic, which is the property of the solid. $t_c$ is larger the viscous behavior starts to appear as the viscosity is proportional to the duration, then the behavior changes to reveal the mixed properties of elasticity and the viscosity. In this case, the subject is not directly the observer but it appears as the impactor which is one part of nature and it contacts with the viscoelastic surface which is another part of nature.  Here the Deborah number (Eq.\ref{eq:E14}) plays the same role of relations to give different intermediate asymptotics for Eq.~\ref{eq:E9} and Eq.~\ref{eq:E12},
 \begin{equation}
 {\rm De}=1/Z  = \frac{\tau}{t_c} =  \frac{\mu v_i}{E \delta_m} = \frac{{\rm relaxation~time} }{{\rm contact~time}}.
\label{eq:E14} 
\end{equation} 

In both cases, the interaction or relation is clearly introduced by a single dimensionless number ${\rm De}$, then it is related to the properties. Here it is found a similar proposition of Rovelli; {\it No interaction, no properties}. In our case, it would be {\it no scale, no properties}.

These cases also show that the phenomena cannot be described from the idea that the observer and nature are distinguishable. It can be described only when both are involved. Thus the description is completed in {\it complicit} manner in which the subject and the object are equally involved. It does not mean that the description is originated from either the subject or the object but it is originated from {\it relation} in which two are fundamentally, indistinguishably involved. This complicity is essential for the relational interpretation on the soft matter. It also concludes that the world cannot be unique in the sense that the world is unique for any subjects or observers. It is because the properties of the pitch and the viscoelastic materials cannot be determined without the interactions or relations between the subjects and the observers. 

This finally leads to one of his conclusions. {\it The world is made of events, not things}\cite{RovEv}. In his interpretation, the difference between {\it things} and {\it events} is {\it that things persist in time; events have a limited duration}\cite{RovEv2}. This paper has shown how the limited duration of the observation of time changes the properties of the pitch. If the limited duration of the observation time is smaller enough than the relaxation time, $T \ll \tau$, the pitch behaves as solid. If the limited duration of the observation time is closer to the relaxation time, $T \sim \tau$, then the pitch starts to reveal the fluid behavior. As for the dynamical contact with viscoelastic boards, the time duration of the contact time plays the same role. If the limited duration of the contact time is much smaller than the relaxation time, $t_c \ll \tau$, the material will respond by the elastic behavior. When the time duration of the contact time is closer to the relaxation time as $t_c \sim \tau$, the response changes and the scaling law also changes. Considering how the time duration changes the nature of the material, we have to conclude that the world is made of events, not things.

From this importance of time-duration, the relational interpretation on soft matter also concludes that {\it facts are relative}. If observers have their own observation time scale, the observational properties can be different from observers. For example, if there is a creature with a life span of 1 ms, water must be solid for that creature, while it is fluid for us. The fact is different depending on the time scale. 

Here we find that the intermediate asymptotics is another version of {\it the way in which one part of nature manifests itself to any other single part of nature}. If its self-similar solution involves the similarity parameters including the scale of the subjet. The intermediate asymptotics which is localized by the limit of its similarity parameter is localized by the subject as well. It means that the subjectivity must essentially constitute to the description, and formalization of the problem though it had been in a hidden way.

\begin{figure}[h!]
\centering
\includegraphics[width=0.5\columnwidth]{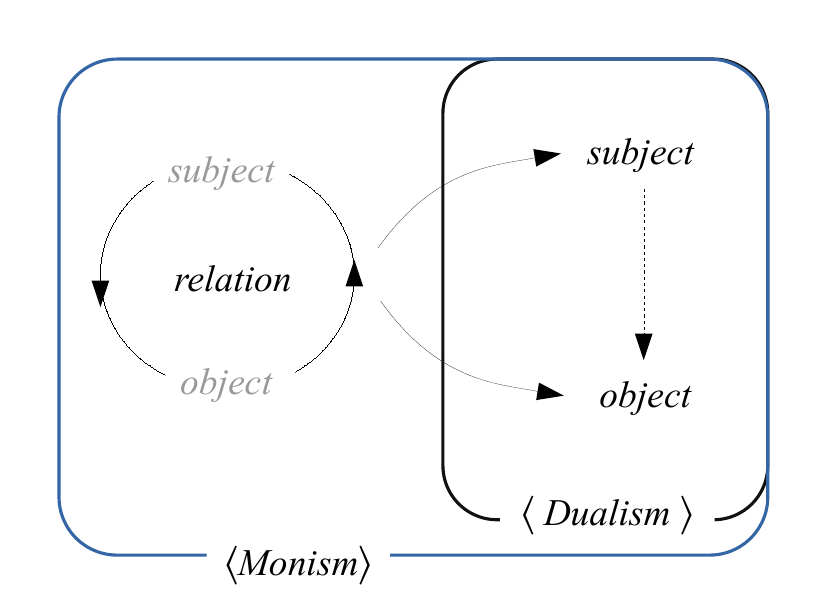}
\caption{Monisitic structure of the relational interpretation. Relations between subjects and objects are origine of the subjects and objects in dual structure.  \label{fig:F2}}
\end{figure}

While we have found that there is an obvious similarity between the Relational interpretation of soft matter and that of quantum mechanics, the two interpretations are expressed in different ways. Relational Quantum Mechanics is formalized in quantum mechanics while the relational interpretation in soft matter is formalized through intermediate asymptotics. Relativity was introduced by the scale-dependence in the relational interpretation of soft matter while Relational Quantum Mechanics introduces relativity through the observer-dependence. However, the fundamental similarity is its monism. The relational interpretation is the monism of relations while the general interpretation is the dualism of subject and object. The dualism is static in the sense that elements have already been distinguished. Monism is dynamic in the sense that the elements are in the process in which they are generated. Relations are the origin of the fundamental dualism of subject and object which constitutes the facts and uniqueness. Uniqueness is realized by this dualism but this dualism is relativized by monism of relations. Relations generate these two elements and can change these two elements. Thus the relational interpretation includes the essence of the general interpretation of physics and that it extends its concept(see Fig.~\ref{fig:F2}). The Relational interpretation reveals the fundamental incompleteness of the process, which relativizes facts and their uniqueness. Such a fundamental structure can be found in both Relational interpretations.   

Dorto also discussed the monistic aspect on the relational quantum mechanics\cite{Dorto}. The context of monism is different though the monism here I define is closer to the {\it priority monism} in which they admit the existence of parts, but holds at the same time that the whole is prior to its parts. In this paper, I discussed monism and dualism in terms of the {\it degree of participation of subjects and objects} for the interpretation\cite{Kemmar}. The author stressed the fundamental participation of subjects to describe the phenomenon based on the intermediate asymptotics. We cannot fundamentally eliminate the participation from subjects as the scale, which corresponds to relations, cannot be determined by the {\it only} objects but determined by the {\it relation} between subjects and objects.  Here we are forced to find the {\it complicit relation} of subjects on the interpretation. These are the major points on the monistic side of the relational interpretation of this paper.

We can say that similar incidents in which relations generate and relativise the duality of subject and object are discussed in other disciplines. Karl von Uexk{\"u}ll described the environment native to organisms as "Umwelt", which is generally translated as the environment in English\cite{Uexkull}.  Organisms possess different organs of perception. When organisms are considered as the unique systems of organs, the environment involving the subject of the organism is fundamentally different, not unique and objective. Such an environment inherent to orgasm was defined as "Umwelt". This concept is also quite similar to relational interpretation in the point that the different relations, which is a different system of orgasm of perception, leads to the fundamentally different environment and it rejects the unique, objective world. Here the subject is fundamentally dependent on the "Umwelt" and two elements are fundamentally entangled. 

A German physician and physiologist, Viktor von Weizs{\"a}cker, also developed a similar idea in his theory of "Gestaltkreis". Weizs{\"a}cker also detected that the subject of a biological system is fundamentally entangled (Verschrankung) with the environment "Umwelt", which can be illustrated as a circular relation. The properties of objects and subjects originate from this circular relation, which corresponds to {\it relations} in the Relational interpretation\cite{Weizsacker}. Bin Kimura, a Japanese psychiatrist, also emphasized the {\it relations}, which is "{\it aida}" in Japanese or should be translated to "{\it betweenness}", to question the condition of the possibility of havinng 'self' naturally established, to explore the mental disease including Schizophrenie\cite{Kimura,Phillips,Fukao}. According to his theory, “{\it something}” is “{\it aida}” or “{\it betweenness},” and it is not that “{\it betweenness}” arises between individuals, but on the contrary, individuals arise on both sides of “{\it betweenness}”\cite{Fukao}. Here we can see the monism of "betweenness" for individuals, which is similar to the monism of relational interpretation. In the relational interpretation, it is not that {\it relations} arises between subjects and objects, but subjects and objects arise on both sides of {\it relations}. As these theories suggest, the relations are expressed by the circular form in which two elements of dualism are entangled. 

Heidegger also criticized the explicitness of the unique, detached, objective world in which we are involved by posing a question to the nature of being. The Relational interpretation is no longer epistemological theory but it is ontological theory in the sense that it does not simply assume any existence without any relation\cite{Dreyfus,Heidegger}. Heidegger's questions and interests are always posed to not {\it being} but {\it the way of being}. The relation is the way how one subject or object manifests itself to others. The Relational interpretation is close to Heidegger's interest and approach.

\section{Conclusion}

In this paper, the fundamental assumption of general physics was characterized and it was demonstrated how the Relational interpretation by Rovelli relativized these assumptions and led to different conclusions. The Relational interpretation localizes the uniqueness by relations. Then I showed that this logical structure can be found in the concept of intermediate asymptotics and how the phenomena of soft matter reveal similar conlusions. Therefore, the phenomenon of soft matter can also be relationally interpreted by the concept of intermediate asymptotics.

The similarity between the two formalizations is thought to derive from the monism of origin. The relational interpretation of quantum mechanics and the relational interpretation of soft matter both begin from the relation. The former starts from the interaction and the latter starts from the scale. Both formalization includes the monistic structure and it relativizes the interpretation in different relations. This structure is also found in other fields.

This work insists that the relational interpretation can be more general and fundamental concept, not a concept for special cases.

\backmatter

\bmhead{Author Contribution Statement}

The author confirms sole responsibility for the following: study conception, design and manuscript
preparation.

\bmhead{Acknowledgments}

The author wishes to thank J. Aames, K. Koseki, M. Murase, A. Koide, Y. Hirai, and the participants of the seminar of Aporia, K. Fukao, K. Shimizu, Y. Miwaki, Y. Maruhashi, Y. Niwa for their fruitful discussion. This work was motivated by the online seminar by the author for the Time and Contingency Research Group, which was organized by R. Ohmaya and I. Motoaki (https://www.youtube.com/watch?v=3mcycHS4xFg).









\begin{thebibliography}{99}
\bibitem{PrigogineOrderI}
See Ref.\cite{PrigogineOrder}. Prigogine \& Stengers summarized the history of development of science, and analyzed how the style of physics is founded. They articulated the point that science has been developed so as that the nature should be dehumanized and considered as the independent being.
\bibitem{PrigogineOrderII}
This two assumptions are also recognised by Prigogine \& Stengers in which they said, “…the experimental dialogue is an irrversible acquisition of human culture. It actually provides a guarantee that when nature is explored by man it is treated as an {\it independent} being.” “Nature not only would be written in a mathematical language that can be desciphered by experimentation, but there would actually exist only one such language. Following this basic conviction, the world is seen as homogenous, and local experimentation can reveal global truth.” See Ref.\cite{PrigogineOrder} p.44.
\end{thebibliography}
\end{document}